\documentclass[preprint]{elsarticle}
\usepackage{amssymb}
\usepackage{graphics}
\usepackage{amsmath}
\usepackage{amsfonts}
\usepackage{bm}% bold math
\usepackage[]{latexsym}
\usepackage{epsfig}
\usepackage{float}
\usepackage{dsfont}
\usepackage{slashed}
\usepackage{xcolor}

\newcommand{\be}{\begin{equation}}
\newcommand{\ee}{\end{equation}}
\newcommand{\bea}{\begin{eqnarray}}
\newcommand{\eea}{\end{eqnarray}}
\newcommand{\brr}{\begin{array}}
\newcommand{\err}{\end{array}}
\newcommand{\bit}{\begin{itemize}}\newcommand{\eit}{\end{itemize}}
\newcommand{\ben}{\begin{enumerate}}\newcommand{\een}{\end{enumerate}}

\newcommand{\ba}{\begin{array}}
\newcommand{\ea}{\end{array}}

\def\lf{\left}

\def\non{\nonumber}\def\pa{\partial}

\def\ri{\right}
\def\al{\alpha}\def\bt{\beta}\def\ga{\gamma}\def\Ga{\Gamma}

\def\la{\lambda}\def\si{\sigma}

\def\1{{_{1}}}\def\2{{_{2}}}

\def\noHe0{:\;\!\!\;\!\!:H_e(0):\;\!\!\;\!\!:}
\def\noHm0{:\;\!\!\;\!\!:H_\mu(0):\;\!\!\;\!\!:}

\def\lf{\left}

\def\non{\nonumber}
\def\pa{\partial}

\def\ri{\right}

\def\al{\alpha}\def\bt{\beta}\def\ga{\gamma}
\def\Ga{\Gamma}

\def\la{\lambda}
\def\si{\sigma}

\def\1{{_{1}}}\def\2{{_{2}}}

\begin{document}
\title{Non-relativistic neutrinos and the weak equivalence principle apparent violation}

\author[a,b]{M.~Blasone}
\ead{blasone@sa.infn.it}

\author[c,d]{P.~Jizba}
\ead{p.jizba@fjfi.cvut.cz}

\author[a,b]{G.~Lambiase}
\ead{lambiase@sa.infn.it}

\author[a,b,c,e]{L.~Petruzziello}
\ead{lupetruzziello@unisa.it}

\address[a]{Dipartimento di Fisica, Universit\`a di Salerno, Via Giovanni Paolo II, 132 84084 Fisciano, Italy}
\address[b]{INFN, Gruppo collegato di Salerno, Italy}
\address[c]{FNSPE, Czech Technical University in Prague, B\v{r}ehov\'{a} 7, 115 19 Praha 1, Czech Republic}
\address[d]{ITP, Freie Universit\"{a}t Berlin, Arnimallee 14, D-14195 Berlin, Germany}
\address[e]{Dipartimento di Ingegneria, Universit\`a di Salerno, Via Giovanni Paolo II, 132 84084 Fisciano, Italy}

%\date{\today}
  \def\be{\begin{equation}}
\def\ee{\end{equation}}
\def\al{\alpha}
\def\bea{\begin{eqnarray}}
\def\eea{\end{eqnarray}}

%\date{\today}%
  \def\be{\begin{equation}}
\def\ee{\end{equation}}
\def\al{\alpha}
\def\bea{\begin{eqnarray}}
\def\eea{\end{eqnarray}}

\begin{abstract}
We study the non-relativistic limit of Dirac equation for mixed neutrinos. We demonstrate that  such a procedure inevitably leads
to a redefinition of the inertial mass. This happens because, in contrast to the case when mixing is absent, the antiparticle sector contribution cannot be neglected for neutrinos with definite flavor. 
We then show that, when a gravitational interaction is switched on, in the weak-field approximation the mass parameter which couples to gravity (gravitational mass) does not undergo the same reformulation as the inertial mass, thus leading to an apparent breakdown of the weak equivalence principle.
\end{abstract}

 \vskip -1.0 truecm
\maketitle

\section{Introduction}
%%%%%%%%%%%%%%%%%%%%%%%%
Neutrino physics has long been considered as an important playground for testing the weak equivalence principle (WEP) both theoretically and experimentally~\cite{wlw,ms,wwg}.
The question of how neutrinos can further strengthen the existing constraints on the equivalence principle or how they can be used to search for its possible violations 
has been discussed in a multitude of papers (i.e. see Refs.~\cite{sck,nmf,sah,ash,bha,ours,ijmpd}), with varying degree of detail and differing conclusions. In addition, over the years several
claims have been made addressing a possible evidence for the incompatibility of WEP with neutrino phenomenology~\cite{gn,hl}.

Virtually, all theoretical studies of neutrino oscillations assume that such particles are
 ultra-relativistic, since typical experimental setups do not allow to deal with non-relativistic
neutrinos. However, this does not mean that the non-relativistic regime is not accessible in principle.
Indeed, with the infusion of new ideas from particle cosmology and astrophysics and the advent of a high precision instrumentation, the behavior of non-relativistic neutrinos has recently been studied from various standpoints. For instance, these particles can exhibit novel features when flavor
oscillations are properly accounted for~\cite{akh}, and their presence can represent a direct evidence for the existence of  cold dark matter~\cite{nie}.
Furthermore, neutrinos that constitute the so-called cosmic neutrino background (CNB) --- also known as relic neutrinos --- may open new scenarios in our understanding of the early
Universe~\cite{wei,fkm}. In fact, it is estimated that the CNB decoupled from matter few seconds after the Big Bang~\cite{rw,gg}.
In this sense, the CNB contains more information on the primordial characteristics of the Universe than the photon-based cosmic microwave background (CMB) radiation.
Since the temperature of the CNB is estimated~\cite{rw} to be  $T\approx 2 K$, it is reasonable to think of relic neutrinos as non-relativistic particles with corresponding virial
velocities of $10^3 - 10^4$km/s. Despite difficulties in detecting these elusive carriers of fundamental knowledge, there are some recent proposals~\cite{ers,gel,lls} that
tend to consider the detection of CNB as a feasible endeavor. In passing, we also want to stress that a finite-temperature analysis is closely linked to WEP violation for quantum systems; for more details, see for instance Refs.~\cite{finitewep}.

In this paper, we study the properties of oscillating neutrinos in the non-relativistic regime. For simplicity, we restrict our analysis to the case of two flavors, thus considering the (coupled) Dirac equations for electron and muon neutrino.
In our investigation, we are partially inspired by Ref.~\cite{zb}, where the authors discussed WEP for various (quantum) particle systems. Though quite general, their discussion does not seem to be directly applicable to oscillating particles such as neutrinos.
{To be more precise, the range of applicability of the usual notion of WEP shall be established with greater care. As a matter of fact, there is an apparent violation of the classical equivalence principle in the case of composite systems with internal degrees of freedom~\cite{zych}, such as for the case of the spin, which non-trivially couples to gravity~\cite{spingrav}. Given that differing and discordant claims can be found in literature which instead tend to preserve WEP even for complex quantum systems~\cite{zych,spinertia}, experimental tests may ultimately have the final word, and several efforts in this direction has already been performed, as shown for instance in Ref.~\cite{exptest}. In this context, it is worth observing that the techniques employed for the aforementioned experiments could in principle be capable of unraveling new macroscopic forces mediated by exotic particles whose strength becomes comparable with the gravitational interaction at certain energy scales~\cite{exptest2}. 
Since flavor neutrinos are regarded as more complicated objects with respect to an elementary particle due to the existence of flavor transitions, it is opportune to keep in mind that the previous observations may be valid also for them.}

In the ultra-relativistic regime, it is widely accepted that the phenomenon of neutrino oscillations is well-described by means of Pontecorvo flavor states~\cite{Pontecorvo}. Here, however, we are interested in the non-relativistic behavior of such states. In this connection, it should be pointed out that, in such a case, corrections to the usual oscillation formula arise when a full-fledged field theoretical approach to neutrino mixing is performed~\cite{blas}. The key aspect is that flavor mixing at the level of fields entails a non-trivial structure at the level of the representation 
(states in the Hilbert space), thus resulting in corrections to the standard Pontecorvo flavor states  
attributable to the rich and complex structure of the flavor vacuum condensate~\cite{blas}.  
Since a quantum-mechanical wave function is a matrix element of the corresponding (quantized) field between the vacuum and a single-particle state,
it imprints information about the vacuum state itself. Indeed, by restricting our analysis of neutrino mixing to non-relativistic quantum mechanics (QM), we find that novel and intriguing effects related to the notion of inertial mass arise in the flavor basis even
without invoking the full quantum field theoretical (QFT) apparatus.

This Letter is organized as follows: in Section~\ref{2}, we study in detail the non-relativistic limit of the Dirac equation for mixed neutrinos and show that in this framework one inevitably comes across a non-trivial correction
to the inertial mass $m_{{\rm{i}} }$.
In addition, if a gravitational field is switched on, we prove in Section~\ref{3} that the ensuing gravitational mass $m_{{\rm{g}} }$
does not undergo the same redefinition as $m_{{\rm{i}} }$, and hence
$m_{{\rm{i}} }\neq m_{{\rm{g}} }$, which is a direct signature of WEP violation.
A brief summary of results and related discussions are
given in Section~\ref{con}. In the Appendix, 
some
finer technical and conceptual details needed in the main text are clarified.
%
%In the end, all the claims revolving around a possible detection of CNB imply an observable evidence of relic neutrinos, which (as already remarked above)
%should be described by means of a non-relativistic limit of known physics.

%%%%%%%%%%%%%%%%%%%%%%%%%%%%%%%%%%%%%%%%%%%%%%%%%%%%%%%%%%%%%%%%%%%%%%
\section{Non-relativistic neutrinos without external field} \label{2}
%%%%%%%%%%%%%%%%%%%%%%%%%%%%%%%%%%%%%%%%%%%%%%%%%%%%%%%%%%%%%%%%%%%%%%
Let us consider the Dirac equation associated with flavor neutrinos $\nu_e$ and $\nu_\mu$. In the simplest case of a two-flavor model and no external field, it reads
\be
\lf(i\gamma^\al\partial_\al \ - \ \mathbb{M}\ri)\Psi \ = \ 0\,.
\label{deq}
\ee
Here, $\gamma^\al$ is implicitly meant to be the $8\times 8$ matrix $\mathbb{I}_{2\times 2}\otimes\gamma^\al$ and
$\mathbb{M}$ is the $8\times 8$ (non-diagonal) mass matrix, which in the $4\times 4$ block formalism reads
\be\label{M}
\mathbb{M} \ = \ \begin{pmatrix}m_e & m_{e\mu} \\m_{e\mu} & m_\mu\end{pmatrix}\!.
\ee
The wave-function $\Psi$ contains the bispinors related both to $\nu_e$ and $\nu_\mu$, i.e.
\be\label{sp}
\Psi=\begin{pmatrix}\psi_e\\\psi_\mu\end{pmatrix}\!.
\ee
If we explicitly write the two Dirac equations, we get
\bea\label{deqe}
\lf(i\ga^\al\partial_\al-m_e\ri)\psi_e&=&m_{e\mu}\psi_\mu\,,\\[2mm]
\label{dequ}
\lf(i\ga^\al\partial_\al-m_\mu\ri)\psi_\mu&=&m_{e\mu}\psi_e\,.
\eea
Unless stated otherwise, we will focus only on Eq.~(\ref{deqe}), since the ensuing results for the muon neutrino are easily obtained by exchanging the subscripts
$e\leftrightarrow\mu$. In addition, with foresight of a non-relativistic treatment of (\ref{deq}) we will employ
the standard Dirac representation of $\gamma$ matrices. Consequently, the
positive-energy wave functions satisfy algebraic equations
\bea\label{e1}\non
\lf(i\partial_0-m_e\ri)\varphi_e \ + \ i{\boldsymbol{\si}}\cdot{\boldsymbol{\nabla}}\chi_e &=& m_{e\mu}\varphi_\mu\,,\\[2mm]
-i{\boldsymbol{\si}}\cdot{\boldsymbol{\nabla}}\varphi_e \ - \ \lf(i\partial_0+m_e\ri)\chi_e &=& m_{e\mu}\chi_\mu\,.
\eea
Here, $\varphi_{e,\mu}$ and $\chi_{e,\mu}$ denote the ``large'' (upper)  and ``small'' (lower) spin components of respective bispinors. At this point, we can perform the non-relativistic limit,
by assuming that the dominant contribution to the energy comes from the rest mass. Hence, in Eqs.~(\ref{e1}) we can assume the kinetic energy to be much smaller than the rest mass. One can thus
pull out from the bispinor the fast oscillating factor $e^{-im_\si t}$ (for the positive energy solutions) so that
\be
\psi_\si (t)\ = \ e^{-im_\si t}\widetilde{\psi}_\si (t)\,, \qquad \si \ = \ \{e,\mu\}\,,
\label{ansa}
\ee
with the field $\widetilde{\psi}_\si$ oscillating much slower than $e^{-im_\si t}$ in time.
Then, one drops the term $\partial_0\widetilde{\psi}_\si$ as small compared to $-2im_{\si}\widetilde{\psi}_\si$ (more specifically, one assumes that  $|i\partial_0\widetilde{\psi}_\si|\ll|2m_\si\widetilde{\psi}_\si|$).
Accordingly, Eqs.~(\ref{e1}) reduce to
\bea\label{e2a} \non
\mbox{\hspace{-4mm}}i\partial_0\widetilde{\varphi}_e \ + \ i{\boldsymbol{\si}}\cdot{\boldsymbol{\nabla}}\ \! \widetilde{\chi}_e &=& m_{e\mu}e^{i\lf(m_e-m_\mu\ri)t}\ \!\widetilde{\varphi}_\mu\,,\\[2mm]\label{e2}
-i{\boldsymbol{\si}}\cdot{\boldsymbol{\nabla}}\ \!\widetilde{\varphi}_e\ - \ 2m_e\widetilde{\chi}_e &=& m_{e\mu}e^{i\lf(m_e-m_\mu\ri)t}\ \!\widetilde{\chi}_\mu\,.
\eea
Analogous relations hold for $\nu_\mu$.
In what follows, we will remove the tilde from the components of Dirac bispinors for simplicity's sake.
Note that, in absence of mixing, the small spin component $\chi$ is negligible with respect to the large one $\varphi$. In presence of mixing and in the non-relativistic limit, however,
the small component $\chi_\mu$ can be of the same order as $\varphi_e$ provided $m_{e\mu}$ is of order $m_{e\mu}\approx |{\boldsymbol{\si}}
\cdot {\boldsymbol{p}}| = |{\boldsymbol{p}}|$. 

Let us now plug $\chi_e$ in the expression for $\varphi_e$. We get
\begin{eqnarray}\label{fe1}
\hspace{-2mm}i\pa_0\ \!\varphi_e &=& -\frac{\nabla^2}{2m_e}\ \! \varphi_e \ + \ e^{i(m_e-m_\mu)t}\lf[m_{e\mu} \varphi_\mu + \frac{i\,m_{e\mu}}{2m_e}\ \! \left({\boldsymbol{\si}}  \cdot {\boldsymbol{\nabla}}\right)\ \! \chi_\mu\ri].
\end{eqnarray}
As expected, the first term on the RHS of Eq.~(\ref{fe1}) represents the kinetic part, whereas the information about the mixing is imprinted in two remaining terms.

One can push the above analysis beyond Eq.~(\ref{fe1}) by employing the ensuing non-relativistic relation
for $\chi_\mu$ stemming from Eq.~(\ref{dequ}). Indeed, by using the fact that
\be\nonumber
\chi_\mu \ = \ -\frac{i\,{\boldsymbol{\si}}  \cdot {\boldsymbol{\nabla}}}{2m_\mu} \ \! \varphi_\mu \ - \ e^{i(m_\mu-m_e)t}\, \frac{m_{e\mu}}{2m_\mu}\ \!\chi_e\,,
\ee
and inserting it into Eq.~(\ref{fe1}), we obtain
\bea\label{fe2}
i\pa_0\ \!\varphi_e =-\frac{\nabla^2}{2m_e}\ \!\varphi_e \ + \ e^{i(m_e-m_\mu)t}\Bigl[m_{e\mu}+\frac{m_{e\mu}}{2m_e}\frac{\nabla^2}{2m_\mu}\Bigr]\varphi_\mu - \frac{i\,m^2_{e\mu}}{4m_em_\mu}\left({\boldsymbol{\si}}  \cdot {\boldsymbol{\nabla}}\right)\ \! \chi_e\, .
\eea
It is clear that we can continue this iteration procedure indefinitely. If the corresponding infinite sum converges, we can get rid of the small spin components
in both $\psi_e$ and $\psi_\mu$ and obtain two coupled field equations for  $\varphi_e$ and $\varphi_\mu$ only --- as it could be expect from the non-relativistic
limit, where only (equal parity) large bispinor components (Pauli spinors) appear.

The aforesaid iterative process brings Eq.~(\ref{fe2}) to the form
\be\label{fe3}
i\pa_0\ \!\varphi_e \ = \ -\,A(\mathbb{M})\frac{\nabla^2}{2m_e}\ \!\varphi_e \ + e^{i(m_e-m_\mu)t} \ B(\mathbb{M})\ \!\varphi_\mu\,,
\ee
where
\be\label{A}
A(\mathbb{M}) \ = \  \sum_{n=0}^\infty\lf(\frac{m_{e\mu}^2}{4m_em_\mu}\ri)^{\!\!n}\,,
\ee
and
\be\label{B}
B(\mathbb{M})\ = \ m_{e\mu}\ + \ \frac{m_{e\mu}}{2m_e}\ \! A(\mathbb{M}) \ \! \frac{\nabla^2}{2m_\mu}\,.
\ee
Since for two flavors the relations between $m_e$, $m_\mu$, $m_{e\mu}$ and the mass
parameters $m_1$ and $m_2$ are known to be\footnote{The relations are obtain by diagonalizing (rotating) the mass matrix ${\mathbb{M}}$ of Eq.~(\ref{deq}).}
\bea\nonumber
m_e&=&m_1\,\mathrm{cos}^2\theta \ + \ m_2\,\mathrm{sin}^2\theta\,,\\[2mm]\nonumber
m_\mu&=&m_1\,\mathrm{sin}^2\theta \ + \ m_2\,\mathrm{cos}^2\theta\,,\\[2mm]
m_{e\mu}&=&\lf(m_2-m_1\ri)\mathrm{sin}\theta\ \!\mathrm{cos}\theta\,,
\label{mass}
\eea
one might easily check that $m^2_{e\mu}<m_em_\mu$. For future convenience, let us denote the expansion parameter $\omega$ as
\be\label{exppar}
\omega=\frac{m_{e\mu}^2}{4m_em_\mu}\,.
\ee
Because $\omega < 1$, the geometric series $A(\mathbb{M})$
converges and it sums up to
\be\label{A2}
A(\mathbb{M}) \ = \ \frac{1}{1-\omega}\,.
\ee
With this, we obtain the equation for the Pauli spinors (large bispinor components) in the Schr\"{o}dinger
form
\bea\label{fe4}
i\partial_0{\varphi}_e=-\lf(\frac{1}{1-\omega}\ri)\frac{\nabla^2}{2m_e}\ \!{\varphi}_e \ + \  \!e^{i\lf(m_e-m_\mu\ri)t}\lf\{m_{e\mu} +\frac{m_{e\mu}\nabla^2}{4m_em_\mu\lf(1-\omega\ri)}\ri\}{\varphi}_\mu\,.
\eea
Equation (\ref{fe4}) is the sought non-relativistic limit of the Dirac equation for an electron neutrino in the presence of mixing.
As already stressed, when we exchange $e\leftrightarrow\mu$ we obtain the corresponding equation for $\varphi_\mu$.

By looking at the formula (\ref{fe4}), we can immediately draw two important conclusions.
First, in order to have a standard kinetic contribution in Eq.~(\ref{fe4}), the would-be inertial mass
$m_e$ should be modified. In fact, we should require that the inertial mass is
$m_e^{{\rm{eff}}}  =   m_e\lf(1-\omega\ri)$.
A similar redefinition must be performed also for $m_\mu$. The existence of $m_e^{{\rm{eff}}} \neq m_e$ might be at first surprising, since it is not evident why mixing
should affect the inertial masses related to flavor states. In this connection, it is worth noting that the presence of the correction term $A(\mathbb{M})$ is due to the
fact that Dirac equation (\ref{e1}) simultaneously deals with large and small bispinor components ($\varphi_e$ and $\chi_\mu$), that in the case of mixing can both be important.
In fact, to reach Eq.~(\ref{fe4}), one has to work interchangeably with small and large components because these are interlocked at all energy scales.
Should the same analysis be performed with the Klein--Gordon equation for mixed fields (i.e. the ones describing mixed composite particles
with spin $0$, such as $K^0$, $D^0$ or $B^0$ mesons~\cite{hara,aba,abu}), an analogous redefinition
of the inertial mass would be found.
We relegate the proof of this latter fact to our future work.

Second, the part related to $\varphi_\mu$ characterizes the oscillation phenomenon. It can be easily checked that the factor inside $\{\ldots \}$ in Eq.~(\ref{fe4}) appears also in the equation for $\varphi_\mu$.
If $\{\ldots \}$ were zero (i.e. when $m_{e\mu} =0$), these two equations would just be two uncoupled equations for free electron and muon neutrinos,
with masses $m_e = m_1$ and $m_{\mu} = m_2$, respectively. However, there is coupling between the two flavor neutrinos by means of the amplitude $\{\ldots \}$, thus implying that there may be
``leakage'' from one flavor to the other. This is nothing but the ``flip-flop'' amplitude of a two-state system~\cite{Feyn}. Note that its modulus is manifestly invariant under the exchange
of flavors $e\leftrightarrow \mu$, which reflects {\em detailed balance} of the oscillation phenomenon.

%%%%%%%%%%%%%%%%%%%%%%%%%%%%%%%%%%%%%%%%%%%%%%%%%%%%%%%%%%%%%%%%%%%%%%
\section{Non-relativistic neutrinos in gravitational field} \label{3}
%%%%%%%%%%%%%%%%%%%%%%%%%%%%%%%%%%%%%%%%%%%%%%%%%%%%%%%%%%%%%%%%%%%%%%

Let us now focus  on what happens  if we switch a gravitational potential on.
It is not a priori evident that the effective inertial masses $m_e^{{\rm{eff}}}$ and $m_\mu^{{\rm{eff}}}$
will also couple to the gravitational potential.
To explore this point, we will restrict our attention on a metric in the
post-Newtonian approximation that goes up to the order $\mathcal{O}\lf(c^{-2}\ri)$. Moreover,
without loss of generality, we will consider the isotropic reference frame, so namely for the gravitational potential we have that $\phi\lf(\vec{x}\ri)\equiv\phi\lf(|\vec{x}|\ri)$.
The ensuing line element reads~\cite{zb}
\be\label{g}
ds^2  =   \lf(1+2\,\phi\,\ri)dt^2 \ -  \lf(1-2\,\phi\ri)\lf(dx^2+dy^2+dz^2\ri)\,.
\ee
In order to couple gravity with the Dirac equation~(\ref{deq}), we use the conventional spin connection formalism. In particular,
we should substitute the slash operator $\slashed{\partial}$ with $\ga^\mu D_\mu$, where $\ga^\mu=e_{\hat{a}}{}^\mu\ga^{\hat{a}}$ and $D_\mu = \partial_\mu+\Ga_\mu$.
$\Ga_\mu$ is the Fock--Kondratenko connection
\begin{eqnarray}
\Ga_\mu \ = \ -\frac{i}{4} \ \!\sigma^{{\hat{a}}{\hat{b}}} \ \!\omega_{\mu{\hat{a}}{\hat{b}}}
\ = \ \frac{1}{8}\lf[\ga^{\hat{a}},\ga^{\hat{b}}\ri]e_{\hat{a}}{}^\la\nabla_\mu e_{{\hat{b}}\la}\,.
\label{fk}
\end{eqnarray}
Here, $\sigma^{{\hat{a}}{\hat{b}}} = i/2 \lf[\ga^{\hat{a}},\ga^{\hat{b}}\ri]$ are the generators of the bi-spinorial representation of Lorentz group, $\omega_{\mu{\hat{a}}{\hat{b}}} = e_{\hat{a}}{}^\la\nabla_\mu e_{{\hat{b}}\la}$ are the spin connection components, $\ga^{\hat{a}}$ represent the
gamma matrices in flat spacetime, $\nabla_\mu$ is the usual
covariant derivative (Levi--Civita connection) and $e_{\hat{a}}{}^\mu$ is the vierbein field.
Note that Latin indices denote the ``Lorentzian'' vierbein labels whereas Greek indices denote manifold coordinate indices.

Because in our case both $g_{\mu\nu}$ and $\eta_{{\hat{a}}{\hat{b}}}$ are diagonal, the evaluation of
the non-vanishing components of the vierbein fields is a simple task. By using the relation
\be\nonumber
g^{\mu\nu} \ = \  e_{\hat{a}}{}^\mu\,e_{\hat{b}}{}^\nu\,\eta^{{\hat{a}}{\hat{b}}}\,,
\ee
we obtain
\be\label{v2}
e_{\hat{0}}{}^0\ = \ 1-\phi\,, \qquad e_{\hat{x}}{}^x \ = \ e_{\hat{y}}{}^y \ = \ e_{\hat{z}}{}^z \ = \ 1+\phi\,,
\ee
and the ensuing Fock--Kondratenko connection
\be\label{fk2}
\Ga_\mu \ = \ \frac{1}{8}\lf[\ga^{\hat{a}},\ga^{\hat{b}}\ri]e_{\hat{a}}{}^\lambda\bigl(\eta_{\mu\lambda}\partial_\rho\phi-\eta_{\mu\rho}\partial_\lambda\phi\bigr)e_{\hat{b}}{}^\rho\,.
\ee
Let us discuss what modifications of Eq.~(\ref{fe4}) will be induced by the
presence of a weak gravitational field. Using the fact that  Eq.~(\ref{deq}) is now replaced by
\be\label{deq2}
\lf(i\gamma^\al D_\al-M\ri)\Psi \ = \ 0\, ,
\ee
we obtain the equations for the electron neutrino sector in the form
\begin{eqnarray}
\mbox{\hspace{-7mm}}\lf(i\pa_0-m_e-i\phi\,\pa_0\ri)\varphi_e+i({\boldsymbol{\si}}  \cdot {\boldsymbol{\nabla}})\ \! \chi_e  &=& m_{e\mu}\varphi_\mu\,,\nonumber \\[2mm]
\mbox{\hspace{-7mm}}-i({\boldsymbol{\si}}  \cdot {\boldsymbol{\nabla}})\ \! \varphi_e-\lf(i\pa_0+m_e-i\phi\,\pa_0\ri)\chi_e &=& m_{e\mu}\chi_\mu\,.
\label{ge}
\end{eqnarray}
The assumption at the basis of Eqs.~(\ref{ge}) is that we consider only a weak gravitation field, i.e. the gravitational potential is slowly varying (as on the Earth surface).
In particular, we consider that $\partial_i\phi\approx 0$, $\forall i$, and so $\phi$ enters in (\ref{ge}) only via vierbeins in $\gamma_{\alpha}$ matrices.

At this point, we can take the non-relativistic limit in Eqs.~(\ref{ge}). This yields
\bea\nonumber
i\pa_0\varphi_e&=&m_e\phi\,\varphi_e  +  e^{i(m_e-m_\mu)t} \ m_{e\mu}\varphi_\mu -  i\left({\boldsymbol{\si}}  \cdot {\boldsymbol{\nabla}}\right)\! \ \chi_e\,,\\[2mm]\label{ge2}
\chi_e&=&-\frac{i\,{\boldsymbol{\si}}  \cdot {\boldsymbol{\nabla}}}{2m_e}\varphi_e  -  e^{i(m_e-m_\mu)t}\,\frac{m_{e\mu}}{2m_e}\chi_\mu\,.
\eea
By following the same procedure which we have
already adopted in the previous Section, one arrives at the non-relativistic
Dirac equation in the presence of a weak gravitational field in the form
\begin{eqnarray}
i\pa_0\varphi_e=\lf(-\frac{\nabla^2}{2m_e^{{\rm{eff}}}}+m_e\,\phi\ri)\varphi_e+e^{i(m_e-m_\mu)t}\lf[\frac{m_{e\mu}}{2m_e}\lf(2m_e+\frac{\nabla^2}{2m_\mu^{{\rm{eff}}}}\ri)\ri]\varphi_\mu\,.
\label{gfe}
\end{eqnarray}
As expected, for the electron neutrino we recover the sum of the kinetic and the potential contribution, but also the same ``flip-flop'' amplitude as in (\ref{fe4}) (at least in the lowest non-trivial weak-field approximation). Notice, however, that whilst the inertial mass undergoes the same redefinition as in the free-field case (\ref{fe4}), the gravitational mass remains $m_e$. This might be seen as a violation of WEP for flavor neutrinos, since $m^{{\rm{eff}}} = m_{{\rm{i}}}\neq m_{{\rm{g}}}$.

It is also interesting to observe that the ``flip-flop'' amplitude can be rewritten as
\begin{eqnarray}
\frac{m_{e\mu}}{2m_e}\lf(2m_e+\frac{\nabla^2}{2m_\mu^{{\rm{eff}}}}\ri) = \ m_{e\mu}\left[1 + \frac{\nabla^2}{2m_e^{{\rm{eff}}}m_\mu^{{\rm{eff}}}\left(1 + \sqrt{1+\frac{m_{e\mu}^2}{m_e^{{\rm{eff}}}m_\mu^{{\rm{eff}}}}}\right)}\right]\! .
\end{eqnarray}
This shows that the amplitude can be entirely formulated in terms of effective inertial masses and $m_{e\mu}$ and, apart from an overall time-dependent phase factor, it is manifestly invariant under exchange of flavors $e\leftrightarrow \mu$ (therefore, it satisfies the detailed balance between $e$ and $\mu$ flavors).

\medskip
Let us finally stress that, should we have performed an analogous treatment in the mass basis, we would
not have found any distinction between inertial and gravitational masses. This holds true because mass eigenstates are  completely decoupled and the absence of off-diagonal mass terms leads to $m_j^{{\rm{eff}}} = m_{j {\rm{i}} } = m_{j {\rm{g}} }$ with $j=1,2$. Consequently, one should be able to retrieve the flavor basis by simply rotating
the state vectors from the mass basis  through the orthogonal transformation
\begin{eqnarray}
 \mbox{\hspace{-4mm}}\left(\begin{array}{c}
   \psi_e \\
   \psi_\mu \\
 \end{array}
  \right)
 = \left(
     \begin{array}{cc}
       \cos \theta & \sin \theta \\
       -\sin \theta & \cos \theta \\
     \end{array}
   \right) \! \left(\begin{array}{c}
   \psi_1 \\
   \psi_2 \\
 \end{array}
  \right) \equiv \mathbb{G} (\theta) \! \left(\begin{array}{c}
   \psi_1 \\
   \psi_2 \\
 \end{array}
  \right)\!,
  \label{III.26.bb}
\end{eqnarray}
which also automatically implies the mass relations~(\ref{mass}). Since Eq.~(\ref{III.26.bb}) should hold at all energies,
it must also be true in the non-relativistic limit. However, it is not difficult to see (cf. Appendix~A) that, given
\begin{eqnarray}
\lf[i(1-\phi)\pa_0-m_1\ri]\varphi_1+i({\boldsymbol{\si}}  \cdot {\boldsymbol{\nabla}})\ \! \chi_1 &=& 0\,,\nonumber \\[2mm]
-i({\boldsymbol{\si}}  \cdot {\boldsymbol{\nabla}})\ \! \varphi_1-\lf[i(1-\phi)\pa_0 + m_1\ri]\chi_1 &=& 0\,,
\label{mbe.a}
\end{eqnarray}
(and an analogous pair of equations for the index $2$), then
\begin{eqnarray}
\mathbb{G}(\theta) \left\{\lim_{\frac{|{\boldsymbol{p}}|}{m_1},\frac{|{\boldsymbol{p}}|}{m_2} \rightarrow 0} \left[\mbox{Eq.~(}\ref{mbe.a}\mbox{)}
 \right] \right\}\,\neq\,\lim_{\frac{|{\boldsymbol{p}}|}{m_e},\frac{|{\boldsymbol{p}}|}{m_\mu} \rightarrow 0} \left[ \mbox{Eq.~(}\ref{ge}\mbox{)} \right] .
 \label{LHSaa}
\end{eqnarray}
In other words, the non-relativistic limit does not commute with the mass-to-flavor rotation.
This apparent contradiction can be easily understood by observing that the non-relativistic limit is implemented by factoring out the fast oscillating phases $e^{-im_{\sigma}t}$ (with $\sigma = \{e,\mu\}$) and $e^{-im_{a}t}$ (with $a = \{1,2\}$). Even though these procedures correctly handle the rest masses in the non-relativistic limit of respective Dirac equations, the corresponding non-relativistic flavor- and mass-basis wave functions $\tilde{\psi}_{\sigma}$ and $\tilde{\psi}_{a}$ (see Eq.~(\ref{ansa})) are not connected via the rotation $\mathbb{G} (\theta)$ anymore.
The actual rotation matrix that operates on the non-relativistic wave functions is more complicated (see Appendix~A) and reduces to $\mathbb{G} (\theta)$ only in the limit when $m_{e\mu} =0$.

%%%%%%%%%%%%%%%%%%%%%%%%%%%%%%%%
\section{Conclusions}\label{con}
%%%%%%%%%%%%%%%%%%%%%%%%%%%%%%%%

In this Letter, we have analyzed the non-relativistic limit of the Dirac equation for mixed neutrinos both in the absence and presence of an external gravitational field.
In its absence, we have shown that
the small components of the flavor bispinor wave functions inevitably
induce a redefinition of the inertial mass.
This rather unexpected behavior is a consequence of the fact that, when mixing is present, in the
Dirac equation one simultaneously deals with large and small bispinor components that are comparably important in the non-relativistic regime.
Furthermore, when an external gravitational field is considered in the weak-field approximation, we have observed that the gravitational mass does not undergo the same redefinition as the inertial one, and hence a violation of WEP arises.
Accordingly, a non-relativistic limit provides a suitable playground for testing the violation of the equivalence principle in neutrino physics, also in light of the novel interpretation which treats such particles as if they were unstable~\cite{teur}, thus validating the exploitation of the results stemming from Ref.~\cite{bonder}. In particular, the latter may become relevant in the context of relic neutrinos in the CNB, which are expected to be detected experimentally in the near future~\cite{Faessler:2017jtg}. 

We recall that the above study has been performed by regarding neutrinos as Dirac fermions. However, we expect that analogous results are also valid for Majorana neutrinos because of the similarity between the two cases in the framework of QFT treatment of mixing and oscillations~\cite{majo}. 

Let us now briefly discuss another conceivable scenario where our analysis might become relevant, namely physics related to sterile neutrinos. To this aim, we define the quantity
\be\label{eqvpar}
\eta=\lf|\frac{m_g}{m_i}-1\ri|\,,
\ee  
which is typically considered in experiments involving WEP violation~\cite{eqprtest}.
In particular, by using the fact that $m_i=m_e(1-\omega)$ and $m_g=m_e$ and invoking that recent experimental bounds on $\eta$ give $\eta\lesssim10^{-11}$, it is straightforward to deduce that the amount of WEP violation in our particular case is quantified by the inequality
\be\label{eqvpar2}
\lf|\frac{\omega}{1-\omega}\ri|\simeq\lf|\omega\ri|=\frac{m_{e\mu}^2}{4m_em_\mu}\lesssim10^{-11}\,.
\ee
With the available sensitivity on $\eta$, the above expression fails to achieve a bound on the absolute value of the neutrino mass better than the one recently obtained with the experiment KATRIN~\cite{katrin}. However, Eq.~(\ref{eqvpar2}) may turn out to be useful in the context of sterile neutrinos~\cite{sterileneutrino}. Indeed, if we focus on a single oscillation channel between sterile right-handed neutrinos and active left-handed neutrinos, it is still possible to adopt the formalism and reasoning employed in this paper. Specifically, if for instance we focus our attention on the electron neutrino disappearance process~\cite{nuedisappear}, Eq.~(\ref{eqvpar2}) can be cast into
\be\label{ster}
\frac{M\sin^22\theta}{4m_1+M\sin^22\theta}\lesssim 4\cdot 10^{-11},
\ee
where $M$ is the mass of the sterile neutrino and where we have made the assumption $M\gg m_1$. Remarkably, the above expression can also be employed in cosmology, since it holds true even for the keV sterile neutrino, which is usually addressed as a potential Dark Matter candidate (cf. see Refs.~\cite{darkmat} for more details).

Now, by resorting to recent data on both light and heavy~\cite{datasterile} sterile neutrinos and by assuming $m_1\approx 1$ eV~\cite{katrin}, we note that
\be\label{ster2}
{M\sin^22\theta}\lesssim 1.6\cdot 10^{-10} \mathrm{eV}.
\ee
This bound is in agreement with the experimental windows available for sterile neutrinos~\cite{datasterile}. Furthermore, it must be highlighted that, should the sensitivity on $\eta$ improve, the constraint arising from Eq.~(\ref{ster2}) may become even stronger than the cosmological ones currently at our disposal. Not surprisingly, our analysis better fits the behavior of right-handed neutrinos due to their heavy mass (if compared with the active ones). 

We want to stress one more time that the results of this paper have been obtained by working in the flavor basis for mixed neutrinos and the simple case of two generations only. We have also discussed how the same procedure is not applicable in the mass basis, because of the non-interchangeability of the  non-relativistic limit and the mixing transformations. 
In this connection, we remark that our analysis supports the view that flavor states correctly describes oscillating neutrinos. This point is of crucial importance in the full-fledged QFT description, because there the choice of either mass or flavor basis corresponds to different unitarily inequivalent
vacuum states~\cite{blas}, which in turn can have observational implications. 
Along this line, we point out that there are also other frameworks in which the above concept becomes relevant. For instance, the study of the inverse $\bt$-decay in accelerated frames has recently shown that general covariance can be fulfilled only when both the Unruh effect and flavor neutrino states are properly taken into 
account~\cite{acc}.

\vspace{2mm}
%%%%%%%%%%%%%%%%%%%%%%%%%%%%
\section*{Acknowledgements}
%%%%%%%%%%%%%%%%%%%%%%%%%%%%
It is a pleasure to acknowledge helpful conversations with
L.~Rachwa{\l} and L.~Smaldone. L.P. is grateful to L. Buoninfante and G.G. Luciano for their support throughout the development of the paper. P.J.  was  supported  by the Czech  Science  Foundation Grant No. 17-33812L.

%%%%%%%%%%%%%%%%%%%%%
\appendix

\section{}
%%%%%%%%%%%%%%%%%%%%

In this appendix we prove the inequality (\ref{LHSaa}). To this end, we concentrate first on the LHS of (\ref{LHSaa}) and, for simplicity,
consider the gravitational potential $\phi$ to be zero. This gives
\begin{eqnarray}
\mbox{\hspace{-24mm}}
\mathbb{G}(\theta)\left(
                    \begin{array}{cccc}
                      i\pa_0 & 0 & 0 & 0 \\
                      0 & -2m_1 & 0 & 0 \\
                      0 & 0 & i\pa_0 & 0 \\
                      0 & 0 & 0 & -2m_2 \\
                    \end{array}
                  \right)\left(
                           \begin{array}{c}
                             \varphi_1 \\
                             \chi_1 \\
                             \varphi_2 \\
                              \chi_2 \\
                           \end{array}
                         \right) =
                         %\nonumber \\[2mm]
%&&\mbox{\hspace{-10mm}}
\mathbb{G}(\theta) \left(
                           \begin{array}{cccc}
                             0 & -i{\boldsymbol{\si}}  \cdot {\boldsymbol{\nabla}} & 0 & 0 \\
                             i{\boldsymbol{\si}}  \cdot {\boldsymbol{\nabla}} & 0 & 0 & 0 \\
                             0 & 0 & 0 & -i{\boldsymbol{\si}}  \cdot {\boldsymbol{\nabla}} \\
                             0 & 0 & i{\boldsymbol{\si}}  \cdot {\boldsymbol{\nabla}} & 0 \\
                           \end{array}
                         \right)\left(
                           \begin{array}{c}
                             \varphi_1 \\
                             \chi_1 \\
                             \varphi_2 \\
                              \chi_2 \\
                           \end{array}
                         \right)
 .
\label{B1}
\end{eqnarray}
By inserting $\mathbb{G}^{-1}(\theta) \mathbb{G}(\theta) = \mathbb{I}$ in front of mass-state bispinors
we can rewrite~(\ref{B1}) as
\begin{eqnarray}
\mbox{\hspace{-18mm}}
\left(
  \begin{array}{cccc}
    i\pa_0 & 0 & 0 & 0 \\
    0 & -2m_e & 0 & -2m_{e\mu} \\
    0 & 0 & i\pa_0 & 0 \\
    0 & -2m_{e\mu} & 0 & -2m_{\mu} \\
  \end{array}
\right) \left(
         \begin{array}{c}
           \varphi_e \\
           \chi_e \\
           \varphi_\mu \\
           \chi_\mu \\
         \end{array}
       \right)
       %\nonumber \\[2mm]
    %&&\hspace{-10mm}
    = \left(
                           \begin{array}{cccc}
                             0 & -i{\boldsymbol{\si}}  \cdot {\boldsymbol{\nabla}} & 0 & 0 \\
                             i{\boldsymbol{\si}}  \cdot {\boldsymbol{\nabla}} & 0 & 0 & 0 \\
                             0 & 0 & 0 & -i{\boldsymbol{\si}}  \cdot {\boldsymbol{\nabla}} \\
                             0 & 0 & i{\boldsymbol{\si}}  \cdot {\boldsymbol{\nabla}} & 0 \\
                           \end{array}
                         \right)\left(
         \begin{array}{c}
           \varphi_e \\
           \chi_e \\
           \varphi_\mu \\
           \chi_\mu \\
         \end{array}
       \right) .
       \label{B2a}
\end{eqnarray}
It is easy to see that the RHS of (\ref{LHSaa}) has the form (again without considering the potential $\phi$)
\begin{eqnarray}
\mbox{\hspace{-27mm}}
\left(
  \begin{array}{cccc}
    i\pa_0 & 0 & - m_{e\mu}e^{\al}& 0 \\
    0 & -2m_e & 0 & -m_{e\mu}e^{\al} \\
    - m_{e\mu}e^{-\al}  & 0 & i\pa_0 & 0 \\
    0 & -m_{e\mu}e^{-\al} & 0 & -2m_{\mu} \\
  \end{array}
\right)\left(
         \begin{array}{c}
           \varphi_e \\
           \chi_e \\
           \varphi_\mu \\
           \chi_\mu \\
         \end{array}
       \right)
   %    \nonumber \\[2mm]
   %&&\mbox{\hspace{-10mm}}
   \ = \ \left(
                           \begin{array}{cccc}
                             0 & -i{\boldsymbol{\si}}  \cdot {\boldsymbol{\nabla}} & 0 & 0 \\
                             i{\boldsymbol{\si}}  \cdot {\boldsymbol{\nabla}} & 0 & 0 & 0 \\
                             0 & 0 & 0 & -i{\boldsymbol{\si}}  \cdot {\boldsymbol{\nabla}} \\
                             0 & 0 & i{\boldsymbol{\si}}  \cdot {\boldsymbol{\nabla}} & 0 \\
                           \end{array}
                         \right)\left(
         \begin{array}{c}
           \varphi_e \\
           \chi_e \\
           \varphi_\mu \\
           \chi_\mu \\
         \end{array}
       \right) ,
       \label{B2aa}
\end{eqnarray}
with $\al=i(m_e-m_\mu)t$.

Clearly, both (\ref{B2a}) and (\ref{B2aa}) are mutually different.
The reason for this discrepancy can be retraced to the fact that the transformation relating the non-relativistic components of the mass and flavor bispinors is not a simple rotation anymore. Indeed, one can easily find that
%the respective wave-function doublets are related as
%
%
\begin{eqnarray}\label{nrr}
         \left(
         %\begin{array}{c}
           \varphi_e,
           \chi_e,
           \varphi_\mu,
           \chi_\mu
         %\end{array}
         \right)^t
				\ = \ \mathbb{\widetilde{G}}(\theta,t)
				         %\lf(\begin{array}{c}
           \left(
           \varphi_1,
           \chi_1,
           \varphi_2,
           \chi_2
         %\end{array}
         \right)^t\,,
\end{eqnarray}
with
\be\label{nrr2}
\mathbb{\widetilde{G}}(\theta,t)   =        \left(
                           \begin{array}{cc}
                   \cos\theta \ \! \exp{\left[i(m_{e\mu}\tan\theta)t\right]} \ \!\mathbb{I}_{2\times 2} & \sin\theta \ \! \exp{\left[-i(m_{e\mu}\cot\theta)t\right]} \ \!\mathbb{I}_{2\times 2}  \\[2mm]
										-\sin\theta \ \! \exp{\left[i(m_{e\mu}\cot\theta)t\right]} \ \!\mathbb{I}_{2\times 2}&  \cos\theta \ \! \exp{\left[-i(m_{e\mu}\tan\theta)t\right]}\ \!\mathbb{I}_{2\times 2} \\
                           \end{array}
                         \right)\,.
\ee

\section*{References}
%%%%%%%%%%%%%%%%%%%%%%%%%%%%%%%%%%%%%%%%%%%%%%%%%%%%%%%%%%%%%%%%%%%%%%%%%%%%%%%%%%%

\end{document}